\documentclass[twocolumn,showpacs,preprintnumbers,amsmath,amssymb]{revtex4}
\usepackage{graphicx}
\usepackage{dcolumn}
\usepackage{bm}

\begin{document}

\preprint{APS}

\title{Itinerant conductance in fuse-antifuse networks}

\author{Cesar I. N. Sampaio Filho$^1\footnote{Correspondence to: cesar@fisica.ufc.br}$, Andr\'{e} A. Moreira$^1$, Nuno A. N. Ara\'{u}jo$^2$, Jos\'{e} S. Andrade Jr.$^{1,3}$, Hans J. Herrmann$^{1,3}$}

\affiliation{$^1$Departamento de F\'{i}sica, Universidade Federal do Cear\'a, 
  60451-970 Fortaleza, Cear\'a, Brasil\\
  $^2$ Departamento  de F\'{i}sica, Faculdade de Ci\^{e}ncias, Universidade de Lisboa, P-1749-016 Lisboa, Portugal, and
  Centro de F\'{i}sica Te\'{o}rica e Computacional, Universidade de Lisboa, P-1749-016 Lisboa, Portugal\\
  $^3$ Computational Physics for Engineering Materials, IfB, ETH Zurich, 
  Schafmattstrasse 6, 8093 Zurich, Switzerland}


\begin{abstract}
We report on a novel dynamic phase in electrical networks, in which current channels perpetually change in time. This occurs when the individual elements of the network are fuse-antifuse devices, namely, become insulators within a certain finite interval of local applied voltages. As a consequence, the macroscopic current exhibits temporal fluctuations which increase with system size. We determine the conditions under which this exotic situation appears by establishing a phase diagram as a function of the applied field and the size of the insulating window. Besides its obvious application as a versatile electronic device, due to its rich variety of behaviors, this network model provides a possible description for particle-laden flow through porous media leading to dynamical clogging and reopening of the local channels in the pore space.
\end{abstract}

                                          
\maketitle

Fluid flow through a porous medium is frequently described in terms of a complex network system of steady-state flow channels, which are more or less tortuous depending on the strength of the disorder \cite{sahimiBook1994,sahimiPhysRep1998}. Preferential channeling  in these systems is a result of minimizing the dissipated energy or flow resistance and is thus typically unique. When the fluid erodes and deposits material, however, the clogging and reopening of channels depends on the evolution of local conditions. We will show here using an electrical analog model that under certain conditions a new itinerant state can be attained in which these preferential channels constantly change their locations.

\begin{figure}[t]
\includegraphics*[width=6.5cm,height=8cm]{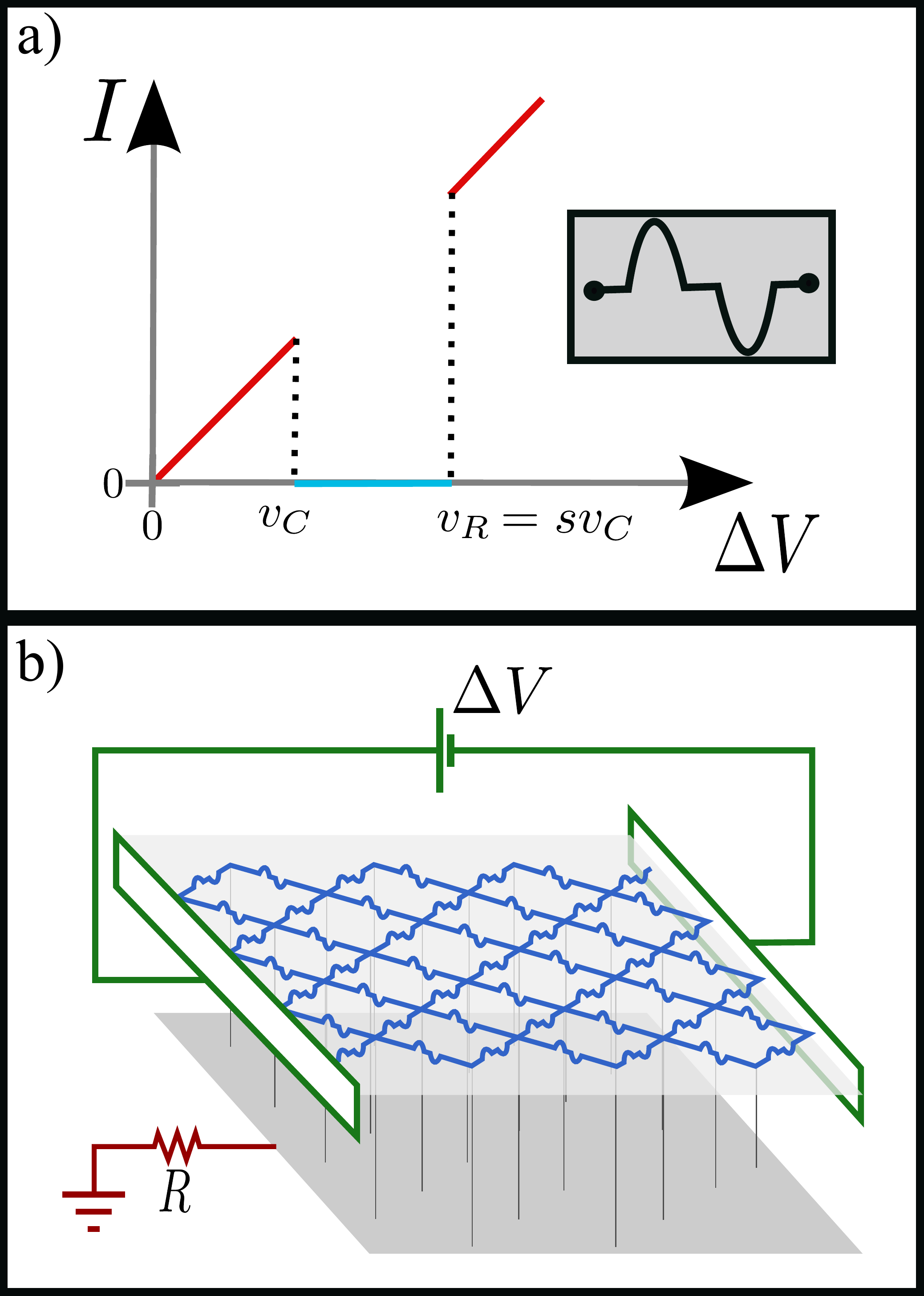}
\caption{(Color online) a) The $I$-$V$ behavior of the fuse-antifuse channel driven by a voltage drop $\Delta V$. If the voltage drop $v$ is below the closing threshold $v_C$, the channel has unitary conductance. If $v_C < v < v_R$, the channel is closed and its conductance is set to zero. Finally, if the voltage drop is above the recovery threshold $v_R$, the channel is again open with unitary conductance. b) Representation of the fuse-antifuse network with each site connected to the ground through a link of high resistance $R\gg 1$.}
\label{fig01}
\end{figure}

A plausible electrical analog for this complex fluid dynamical system is a network in which each link contains a reversible fuse-antifuse device \cite{antifuse1988,antifuse1991,antifuse1993,antifuse1993B,hauckIEEE1998}, as described in Fig.~\ref{fig01}a. In an electrical circuit, a typical fuse behaves as a conductor if the voltage drop is below a given threshold $v_C$, and becomes irreversibly an insulator otherwise. In the case of an antifuse, if the voltage drop exceeds a threshold $v_R$, its behavior changes abruptly  from an insulator to a conductor. This type of switch has been  widely used in programmable elements, in order to configure logic circuits and create customized designs \cite{antifuse1988,antifuse1991}. As depicted in Fig.~\ref{fig01}b, more sophisticated devices like programmable read-only memories \cite{antifuse1993}, can intrinsically couple in their bits both fuse and antifuse behaviors in a reversible fashion, which are triggered one or another depending on the applied potential drop \cite{antifuse1993,damageBook2000,hansenRP2010}.

Our model for particulate transport in a fluid flowing through a clogging-reopening pore space consists of a regular lattice where each link is a reversible fuse-antifuse device (see Fig.~\ref{fig01}). These links have the same conductance $g=1$ while in the conducting state, namely for $v<v_C$ and $v>v_R$, but are associated to randomly distributed closing and recovering threshold voltages, $v_C$  and $v_R$, respectively  \cite{hansJPL1985,hansPRB1988,zapperiEPJB2000,zapperi2006,andradeEPhys2012,moreiraPRL2012,hansPRE2012,oliveiraPRL2014,moreiraPRL2014}. The values of $v_{C}$ are chosen from a uniform distribution in the interval $\left[\epsilon, \epsilon + \Delta \right]$, with $\epsilon = 0.10$ and $\Delta = 0.1$ or $\Delta = 1.0$. The recovering threshold is defined as $v_{R} = sv_{C}$, where $s$ is a parameter describing how much the closing and opening events are separated. Moreover, all sites of the lattice, except those at which the overall voltage drop is applied, are connected to the ground through a link of very high resistivity, $R\gg 1$. Therefore, sites that will become eventually isolated during the dynamics will have zero potential.

\begin{figure}[h]
\includegraphics*[width=\columnwidth]{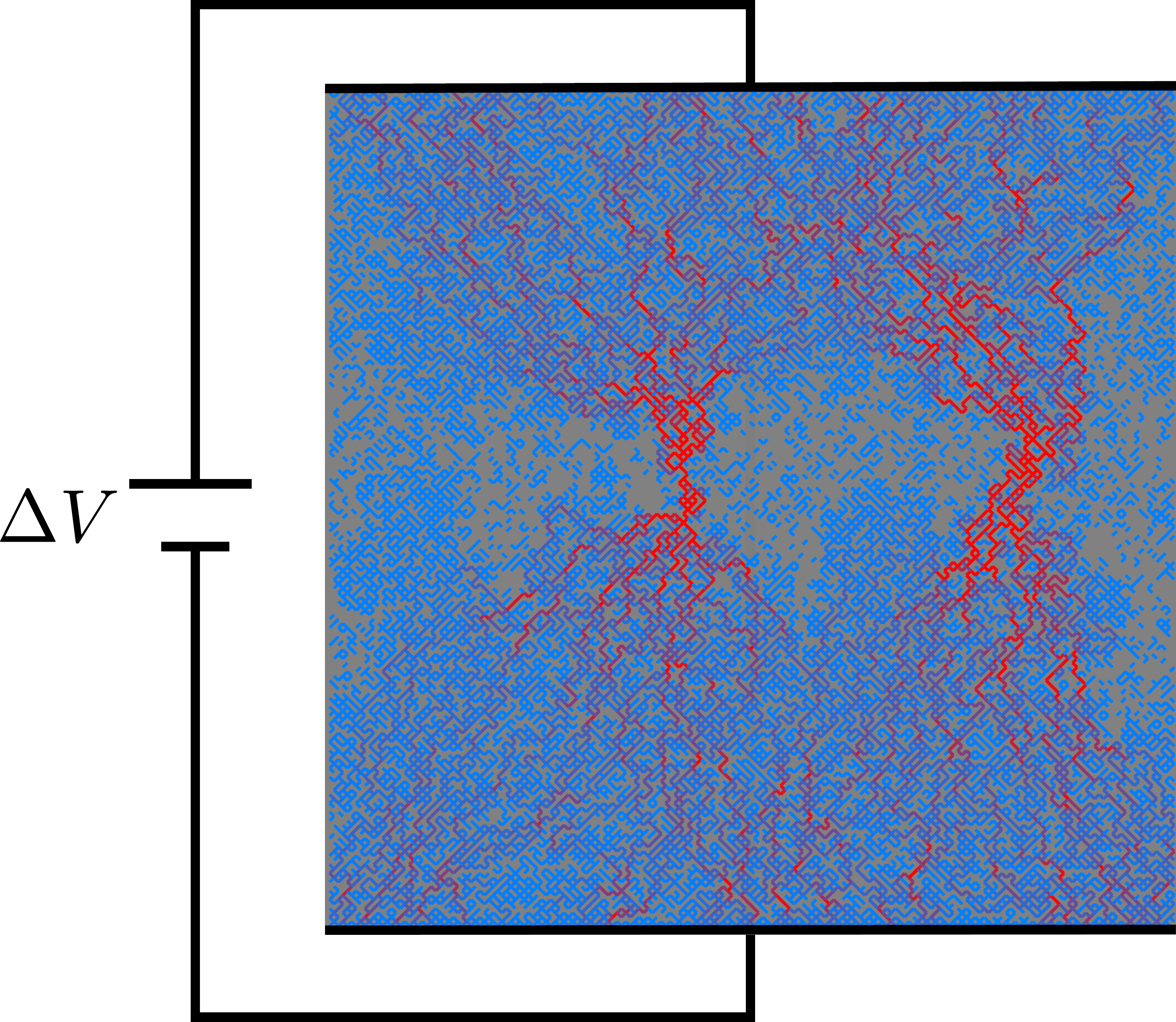}
\caption{(Color online) Typical snapshot of the itinerant phase of the system at steady state, for a tilted square lattice with $\Delta = 0.1$, $sL^{-1} = 4$, $VL^{-1}=2$, and $L=96$. A potential drop $\Delta V$ is applied from top to bottom and periodic boundary conditions from left to right. Closed bonds are represented in brown while open bonds are colored according to the their local current intensity, from blue to red.}
\label{fig02}
\end{figure}

We performed simulations with an initially filled tilted square lattice and apply a voltage drop $V$ across it. Next, at each step, we solve Kirchhoff's law \footnote{The resulting system of linear algebraic equation is solved through the HSL library, a collection of FORTRAN codes for large-scale scientific computation. See http://www.hsl.rl.ac.uk/.} on each site in order to determine  the voltage drop $v_{k}$ on each bond $k$. After determining $v_{k}$, we proceed to identify  which bonds will be closed or reopened, by calculating for each bond,
\begin{equation}
R_{k} = \frac{\vert v_{k} \vert}{v_{C}}.
\label{eq01}
\end{equation}
Any bond $k$ with $1 < R_{k} < s$ will be removed, that is, will have its conductivity set to zero in the following step. Note that this removal is reversible, that is, if the voltage drop between the endpoints of a removed bond changes and $R_{k}$ becomes larger than $s$, the bond will be recovered to the system with original conductivity. Since the removal/recovery process is synchronous, that is, several bonds may change state at the same step, it is possible that entire regions of the system disconnect from both terminals. In this case the potential in these regions becomes zero. Here a time unit is defined in terms of a model iteration in which  Kirchhoff's law is simultaneously solved for all sites. After the corresponding voltage drops are computed, we can further identify which ones remain closed or open, or will be closed or reopened.

A snapshot of the system at the steady state of the itinerant phase is shown in Fig.~2, where the bonds represent conducting channels colored according to their associated values of electrical currents. In order to quantify the macroscopic behavior of the system, we calculate the global conductance $G$, which is defined as the ratio between $\sum_k v_{k}g_{k}$ and the global voltage drop applied to the system $V$, where the summation stands for all fuse-antifuse elements along any line perpendicular to the direction where $V$ is applied, and $g_{k}$ is the conductance of the $k$-th bond, here considered unitary. 

\begin{figure}[t]
\includegraphics*[width=\columnwidth]{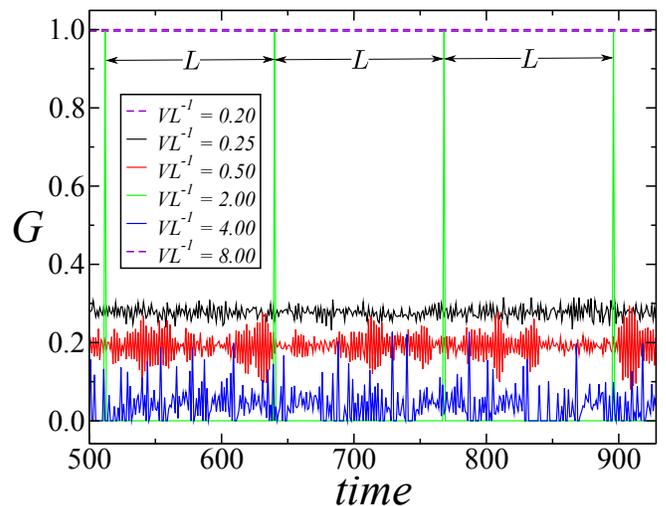}
\caption{(Color online) The time dependence of the global conductance $G$. By fixing the values of $\Delta=0.10$, $sL^{-1}=0.10$, and $L=128$, for sufficiently low ($VL^{-1}=0.20$) and high ($VL^{-1}=8.0$) values of the voltage drop, all elements of the network remain in the conducting state. For $VL^{-1}=0.25$ and $0.50$, the conductance is always finite, $G>0$, but fluctuates around mean values. For $VL^{-1}=2.0$, $G$ exhibits periodic peaks of maximal conductance at a frequency that is proportional to $L^{-1}$. Finally, for a larger voltage drop, $VL^{-1} = 4.0$, fluctuations are again observed in $G$, but now with episodes of insulator behavior, $G=0$, due to eventual losses of global network connectivity.}
\label{fig03}
\end{figure} 

Next, we consider the dependence of the global conductance $G$ on the parameter space, defined in terms of the scaled variables $VL^{-1} \times sL^{-1}$. As shown in Fig.~\ref{fig03}, the time dependence of the conductance $G$ can exhibit a rich variety of behaviors, depending on the applied scaled voltage drop $VL^{-1}$. By fixing the values of $\Delta=0.10$, $sL^{-1}=0.10$ and $L=128$, for sufficiently low ($VL^{-1} = 0.2$) and high ($VL^{-1} = 8.0$) values of the voltage drop,  since all elements of the  network remain in the conducting state, the system operates under an invariant and maximal conductance, $G_{max}=1$. For $VL^{-1} = 0.25$ and $0.50$, the fuse-antifuse elements can dynamically close and open in the network. As a consequence, after a transient period, the conductance is always finite, $G>0$, but fluctuates around an approximately constant mean value. For $VL^{-1}=2.0$, $G$ exhibits periodic peaks of maximal conductance at a frequency that is proportional to $L^{-1}$. Finally, for a larger voltage drop, $VL^{-1} = 4.0$, fluctuations are again observed in $G$, but now with episodes of insulator behavior, $G=0$, due to eventual losses of global connectivity. 

\begin{figure}[t]
\includegraphics*[width=\columnwidth]{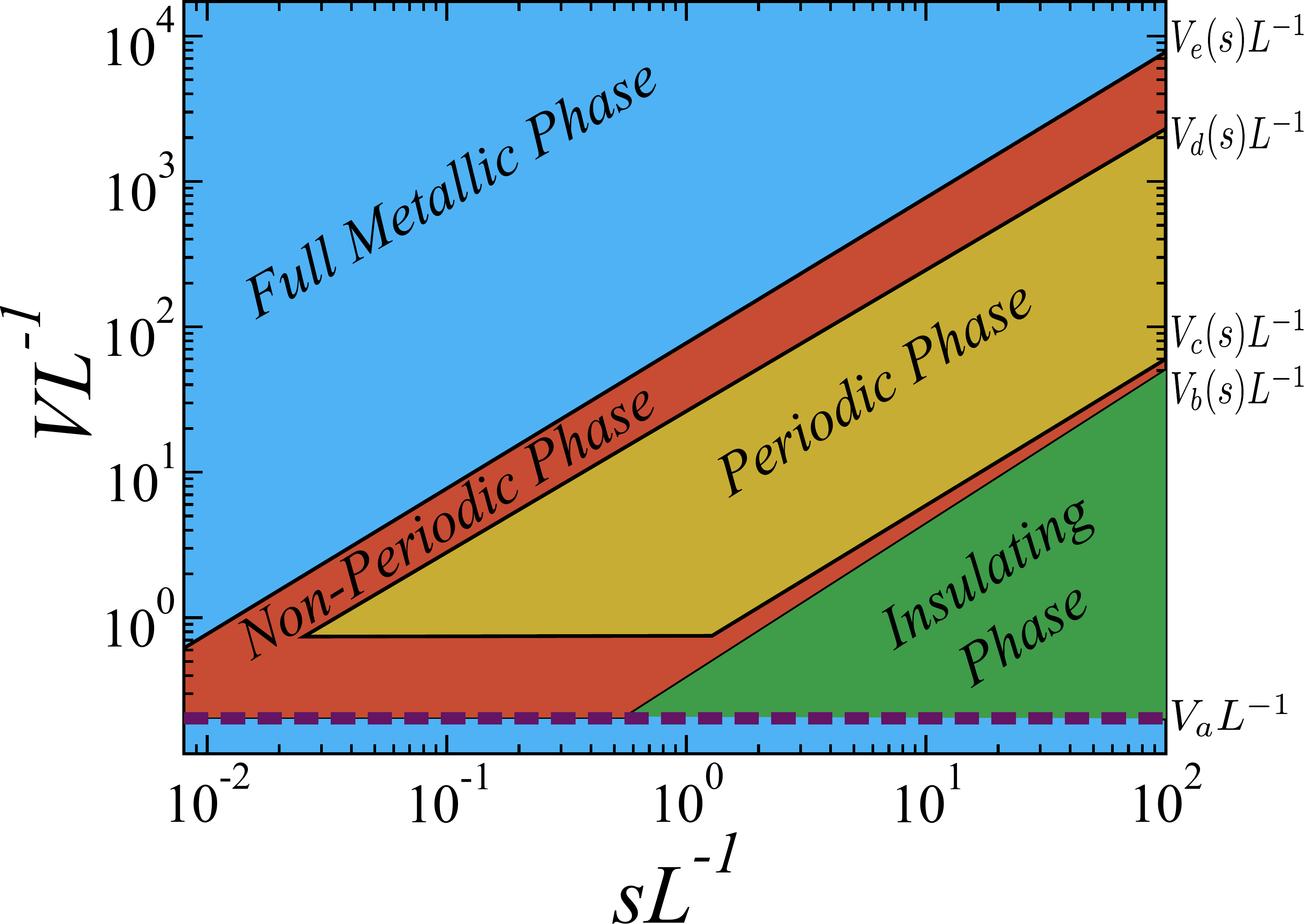}
\caption{(Color online) The phase diagram of the fuse-antifuse networks for $\Delta = 0.10$. The diagram is presented in terms of the scaled voltage drop $VL^{-1}$ and the scaled parameter $sL^{-1}$. Four phases can be identified. If the voltage drop is sufficently small ($V < V_{a}$) or sufficiently high ($V>~V_e(s)$), the system is in the \textit{full metallic phase} (blue). If the voltage drop is $V_a \leq V < V_{b}(s)$, the system is in the \textit{insulating phase} (green). When the voltage drop is in the range $V_b(s) \leq V < V_{e}(s)$, the system is in the \textit{itinerant phase}, which is divided in periodic and non-periodic regimes. The periodic regime is bounded by the potentials $V_{c}(s)$ and $V_{d}(s)$. All voltages $V_a, V_b, V_c, V_d$, and $V_e$ grow linearly with system size.}
\label{fig04}
\end{figure}

\begin{figure}[t]
\includegraphics*[width=\columnwidth]{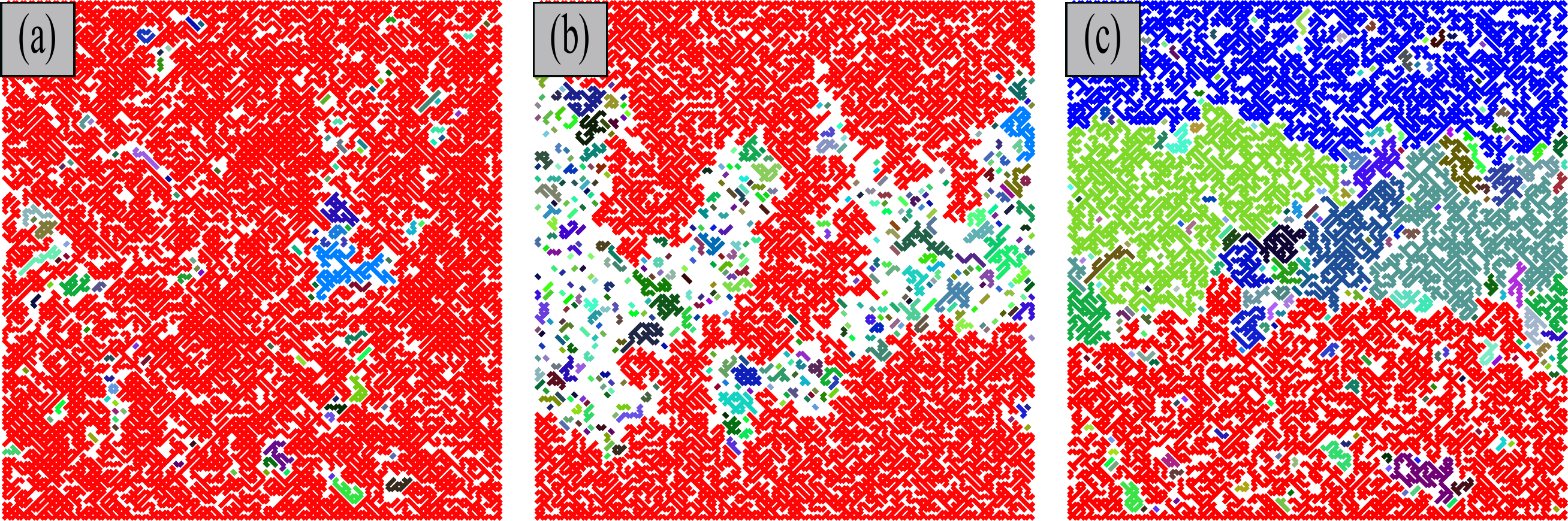}
\caption{(Color online) Snapshots for different phases of a system with $L=64$. The largest cluster is represented in red. a) The non-periodic itinerant phase (below the periodic regime) with $VL^{-1}=0.5$ and $sL^{-1}=0.1$, where the system is highly connected. b) The non-periodic itinerant phase (above the periodic regime) with $VL^{-1}=2.0$ and $sL^{-1}=0.1$ where the system is lower connected. c) The insulating phase, where the largest cluster does not percolate along the direction where the voltage drop is applied.}
\label{fig05} 
\end{figure} 

In Fig.~\ref{fig04} we show the phase diagram for the case $\Delta = 0.10$, in terms of the scaled parameters $VL^{-1}$ and $sL^{-1}$. Four phases can be clearly  identified. If the voltage drop is sufficiently small ($V < V_{a}$) or sufficiently high ($V>~V_e(s)$), all connections conduct and the system preserves its maximal conductance. This behavior characterizes the \textit{full metallic phase}. Moreover, the voltage drop $V_{a}L^{-1}$ is independent on $s$, depending only on the distribution of thresholds. Next, if the voltage drop is in the range $V_a < V < V_{b}(s)$, the system is in the \textit{insulating phase} (green in Fig.~\ref{fig04}), where the global conductance at steady state is zero, since a spanning percolation cluster of conducting fuse-antifuse elements is not present. When the voltage drop is in the range $V_b(s) < V < V_{e}(s)$, opening and burning events occur and we observe fluctuations in the global conductance. This characterizes the \textit{itinerant phase}, which is divided in \textit{periodic} (yellow) and \textit{non-periodic regimes} (brown). The periodic regime is bounded by the potentials $V_{c}(s)$ and $V_{d}(s)$, and the global conductance oscillates between zero and its maximal value, $G_{max}=1.0$, with frequency equals to $L^{-1}$. All voltages $V_a, V_b, V_c, V_d$, and $V_e$ increase linearly with system size. In Fig.~\ref{fig05} we show snapshots of steady-state configurations for three regions of the phase diagram. While for the non-periodic itinerant phase the system percolates along the direction where the voltage drop is applied, with different levels of connectivity (see Figs.~\ref{fig05}a and \ref{fig05}b), for the insulating phase the largest cluster does not percolate (see Fig.~\ref{fig05}c). 

\begin{figure}[t]
\includegraphics*[width=\columnwidth]{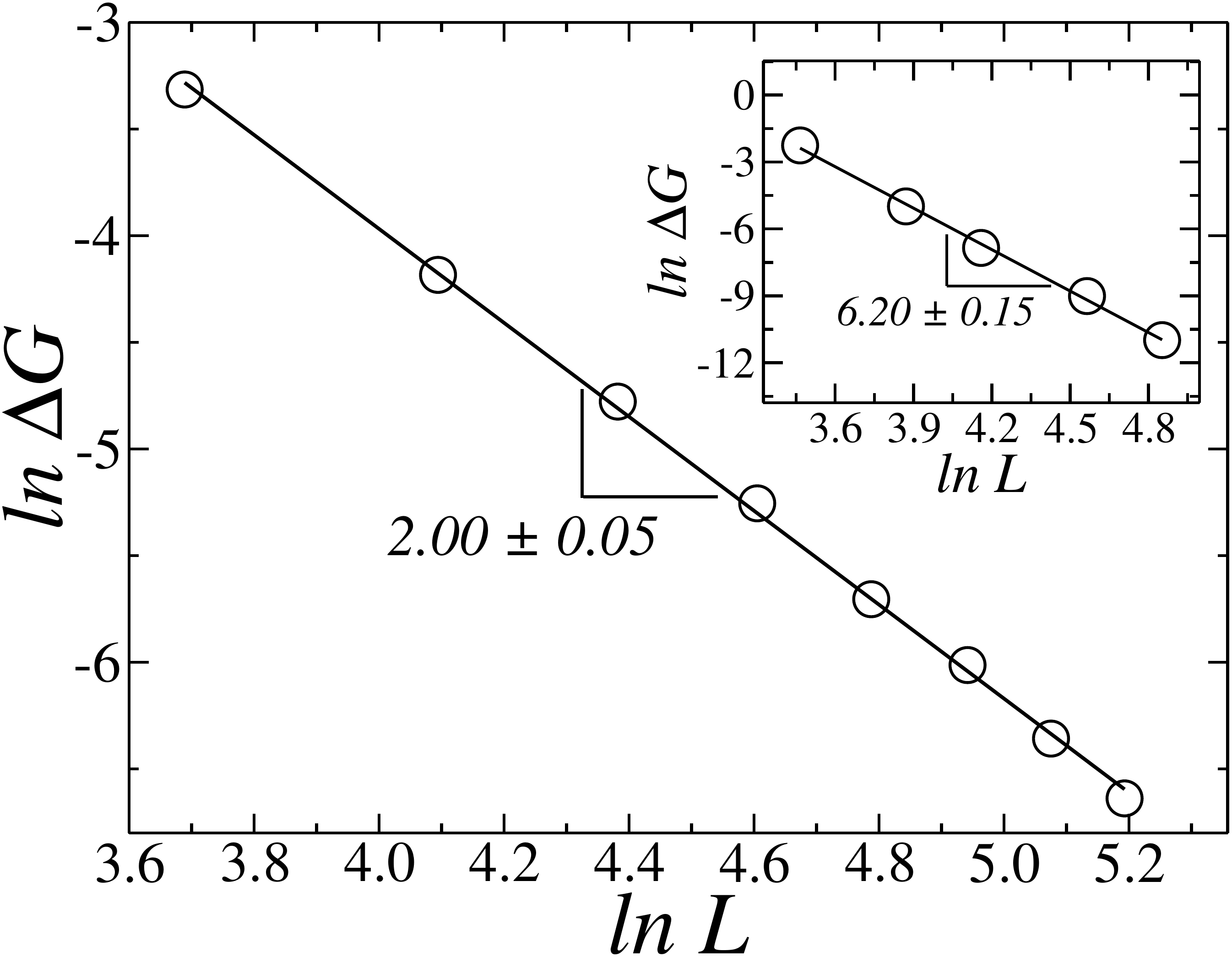}
\caption{Logarithmic plot showing the size dependence of the relative fluctuation $\Delta G$ of the current for $\Delta = 0.10$ and $s = 12$. For each data point we consider $10^5$ time steps at steady state averaged over $100$ samples. The least-squares fit to the data of a power law, $\Delta G~\sim L^{-\theta}$, gives the exponent $\theta = 2.00\pm 0.05$, corresponding to a strongly self-averaging behavior \cite{binderJStat1991}. For $\Delta = 1.0$, we obtain $\theta = 6.20\pm 0.15$.}
\label{fig06}
\end{figure}

\begin{figure}[t]
\includegraphics*[width=\columnwidth]{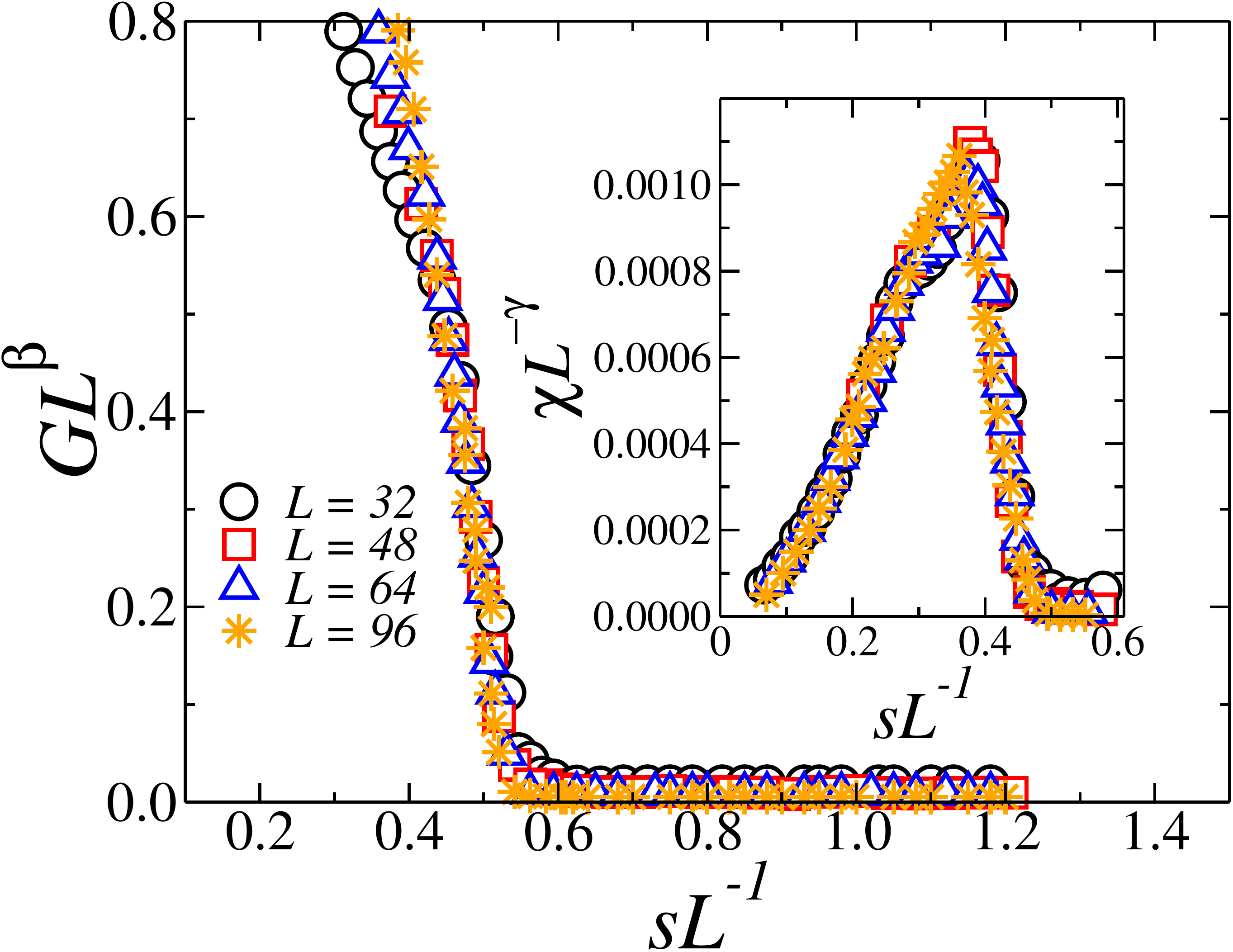}
\caption{(Color online) Data collapse of the mean conductance for different lattice sizes $L$ as a function of the parameter $sL^{-1}$, considering $\Delta=0.10$ and $L=32$ (circles), $48$ (rectangles), $64$ (triangles), and $96$ (stars). Each data point is measured in the steady state during $10^5$ time steps and averaged over $100$ samples. The inset shows the data collapse of the weighted variance, defined by $\chi_{G} = N\left(\left\langle G^{2}\right\rangle-\left\langle G\right\rangle^{2}\right)$, where $N = 2L^{2}$ is the number of sites on the lattice. The best data collapse is obtained for $\beta = 0.58$ and $\gamma = 1.70$. Both results suggest that the fuse-antifuse model undergoes a continuous phase transformation from a conductive to an insulating phase.}
\label{fig07}
\end{figure}

In order to analyse the steady state of the itinerant phase, we measure the relative fluctuation of $G$,
\begin{equation}
\Delta G = \frac{\left\langle G^{2} \right\rangle - \left\langle G \right\rangle^{2}}{\left\langle G\right\rangle^{2}},
\label{eq02}
\end{equation}    
which scales with the system size as, $\Delta G~\sim L^{-\theta}$, where the exponent $\delta$ determines the degree of self-averaging. If $\theta = d$, with $d$ being the dimension of the lattice, the quantity measured is strongly self-averaging. If $0 <\theta< d$, the quantity is self-averaging, and if $\theta = 0$ the quantity lacks self-averaging \cite{binderJStat1991,harrisPRL1996,domanyPRE1998,moghaddamJPhys2002,dsouzaPRL2014,SaberiPhysRep2015,orlandini2002,olivePRB2008,novikovPNAS2014,campbellPRE2016}. As shown in Fig.~\ref{fig06}, the least-squares fits of power laws to the data sets of $\Delta G$ against  $L$ results in $\theta=2.00\pm 0.05$ and $6.20\pm 0.15$ for $\Delta=0.1$ and $1.0$, respectively. 

Along the line $V=V_a$ of the phase diagram, as we increase the parameter $sL^{-1}$, the system undergoes a phase transformation from the itinerant (non-periodic) to the insulating phase (the horizontal dashed line in Fig.~\ref{fig04}). In order to characterize this transformation, we analyse the global conductance and its fluctuations near the transition point between the phases, by fixing $V=V_a$ and changing the parameter $sL^{-1}$. In the main plot of Fig.~\ref{fig07}, we show the data collapse of the global conductance as a function of $sL^{-1}$, considering $L = 32, 48, 64$, and  $96$. Each data point is measured at steady state during $10^ 5$ time steps and averaged over $100$ samples. The best data collapse is obtained using the exponent $\beta = 0.58$. The results shown in Fig.~\ref{fig07} suggest that the fuse-antifuse model undergoes a continuous transition from an itinerant to an insulating phase, at a specific scale invariant value $s_{c}L^{-1}$. This is corroborated by the results presented in the inset of Fig.~\ref{fig07}, where we show the data collapse of the weighted variance $\chi_G$, with $\chi_{G} = N\left(\left\langle G^{2} \right\rangle - \left\langle G \right\rangle^{2}\right)$, using the exponent $\gamma = 1.70$.

In summary, we have introduced a fuse-antifuse network model which exhibits a regime of itinerant conductivity. At fixed external potential, the conductance spontaneously fluctuates with an amplitude that increases with system size. This novel itinerant regime is in some sense similar to the dynamics of braiding rivers \cite{murrayNature1994} and might help to understand flicker noise \cite{duttaRevModPhys1981,pankajPRL2014,paladinoRevModPhys2014} and similar phenomena where spontaneous macroscopic fluctuations appear. Concerning the eroding and depositing fluid flowing through a porous medium, very recent experiments do indeed seem to show, under certain conditions, intermittently fluctuating permeabilities \footnote{F. Bianchi, private communication}.  

\begin{acknowledgments}
We thank the Brazilian agencies CNPq, CAPES, FUNCAP, the National Institute of Science and Technology for Complex Systems, and European Research Council (ERC) Advanced Grant 319968 FlowCCS for financial support. N.A. acknowledges financial support from the Portuguese Foundation for Science and Technology (FCT) under Contracts nos. EXCL/FIS-NAN/0083/2012, UID/FIS/00618/2013, and IF/00255/2013.
\end{acknowledgments}

\end{document}